\documentclass{article}
\usepackage{amsfonts}
\usepackage{amssymb}
\usepackage{amsmath}
\usepackage{amsthm}
\usepackage{mathtools}
\begin{document}

\title{Gravitational Waves in New General Relativity}

\author{Alexey Golovnev$^1$, A. N. Semenova$^2$, V.P. Vandeev$^2$\\
{\small \it $^1$Centre for Theoretical Physics, The British University in Egypt,}\\
{\small \it El Sherouk City, Cairo 11837, Egypt}\\
{\small  agolovnev@yandex.ru}\\
{\small \it $^2$Petersburg Nuclear Physics Institute of National Research Centre ``Kurchatov Institute'',}\\ 
{\small \it Gatchina, 188300, Russia}\\
{\small ala.semenova@gmail.com \hspace{50 pt} vandeev{\_}vp@pnpi.nrcki.ru} } 
\date{}

\maketitle

\begin{abstract}

The models of New General Relativity have recently got attention of  research community, and there are some works studying their dynamical properties. The formal aspects of this investigation have been mostly restricted to the primary constraints in the Hamiltonian analysis. However, it is by far not enough for counting their degrees of freedom or judging whether they are any good and viable. In this paper we study linearised equations in vacuum around the trivial Minkowski tetrad. By taking the approach of cosmological perturbation theory we show that the numbers of primary constraints are very easily seen without any need of genuine Hamiltonian techniques, and give the full count of linearised degrees of freedom in the weak field limit of each and every version of New General Relativity without matter.

\end{abstract}

\section{Introduction}

In the last decade or so, the teleparallel models \cite{revtor} have been very popular for modifying gravity together with looking at their cosmological and astrophysical applications. The most widely used of them are theories of $f(\mathbb T)$ type which do have many interesting features but at the same time are not free of rather serious flaws in their foundations \cite{issues}, to say the least. It invites us to also try another option which has been around for a long time, namely the New General Relativity \cite{HaSh}, or New GR for short.

In (metric) teleparallel models, the dynamical variable is a co-tetrad $e^a_{\mu}$, a set of four linearly-independent differential 1-forms, which are orthonormal by the very definition of the model, that is the metric is {\it defined} as
\begin{equation}
\label{themetr}
g_{\mu\nu}=\eta_{ab} e^a_{\mu} e^b_{\nu}
\end{equation}
with $\eta_{ab}=\mathrm{diag} (+1, -1, -1, -1)$ being the Minkowski tensor. The tetrad $e^{\mu}_a$ is an orthonormal set of vectors defined as a dual basis to, or a matrix inverse of the co-tetrad; and the teleparallel connection
\begin{equation}
\label{theconn}
\Gamma^{\alpha}_{\mu\nu}= e^{\alpha}_a \partial_{\mu}e^a_{\nu} 
\end{equation}
 is then both curvature-free and non-metricity-free. To avoid any confusion, let us mention that we follow the pure tetrad approach to modified teleparallel gravity, that is the spin connection is put to zero from the very beginning. There are many discussions in the literature on these topics \cite{others, discuss}. We share the viewpoint that the pure tetrad approach is the most natural one \cite{meGeom, Lor}, though in many cases using a non-zero spin connection might be beneficial \cite{goodcov}. And in any case, the pure-tetrad description is a legitimate gauge choice \cite{meCov} in the fully covariant approach to teleparalel gravity.

The Teleparallel Equivalent of General Relativity (TEGR) is defined in terms of the basic torsion quantities
\begin{equation}
\label{torstuff}
T^{\alpha}_{\hphantom{\alpha}\mu\nu}=\Gamma^{\alpha}_{\mu\nu} - \Gamma^{\alpha}_{\nu\mu}, \qquad T_{\mu}=T^{\alpha}_{\hphantom{\alpha}\mu\alpha}, \qquad {\mathbb T}=\frac12 T_{\alpha\mu\nu} S^{\alpha\mu\nu}=\frac14 T_{\alpha\mu\nu}T^{\alpha\mu\nu} + \frac12 T_{\alpha\mu\nu}T^{\mu\alpha\nu} - T_{\mu}T^{\mu} 
\end{equation}
as a theory with an action $S=\int d^4 \sqrt{-g}\cdot \mathbb T$ which is fully equivalent to General Relativity, for its Lagrangian density is different from minus the scalar Riemannian curvature of the metric $g_{\mu\nu}$ by only a total covariant divergence term. The non-linear models substitute the torsion  scalar $\mathbb T$ by a non-linear function of it, $f(\mathbb T)$, while the New General Relativity generalises the scalar itself:
\begin{equation}
\label{torsionscalar}
{\mathfrak T}=\frac12 T_{\alpha\mu\nu} {\mathfrak S}^{\alpha\mu\nu}, \qquad {\mathfrak S}_{\alpha\mu\nu} = \frac{a}{2} T_{\alpha\mu\nu} +\frac{b}{2}\left(T_{\mu\alpha\nu} -T_{\nu\alpha\mu} \right)+c\left( g_{\alpha\mu}T_{\nu} - g_{\alpha\nu}T_{\mu}\right)
\end{equation}
with the case of $a=b=c=1$ giving the scalar $\mathbb T$ of TEGR=GR and $f(\mathbb T)$.

Starting from the New GR action
$$S=\int d^4 x \sqrt{-g} \cdot \mathfrak T,$$
the primary constraints of these models have been studied \cite{Dan1, Shy}. Of course, it is a very important message, to start with. However, it is not yet the full dynamical information since there might be, and normally there actually are, secondary constraints, too. At the very least, there should be some traces of the diffeomorphism gauge symmetry "hitting twice". In this paper, we give a thorough analysis of dynamical modes in vacuum around the trivial Minkowski-space co-tetrad $e^a_{\mu}=\delta^a_{\mu}$, not restricting ourselves to the principal symbol only, unlike in Ref. \cite{GWtel}. Of course, our analysis isn't a full answer either. For example, the $f(\mathbb T)$ gravity has many problems of strong coupling \cite{issues}, and precisely the Minkowski spacetime is the most obvious and immediate example of that. What is even more astonishing is that it persists in cosmology, too \cite{meTo}. However, the case of New GR is very much different in this respect from $f(\mathbb T)$, and we think that it is a necessary step if we are to take these models seriously.

The plan of this paper is as follows. In the next Section we review the basic properties of the New General Relativity equations of motion, construct its weak field limit, and discuss the gauge freedom of diffeomorphisms. Then Sections 3, 4, and 5 present all the equations in the tensor, scalar, and vector sectors respectively. In Section 6 we compare our results with the previous works on primary constraints and describe all the nine non-trivial classes of these models. Finally, in Section 7 we conclude.

\section{Dynamics of New General Relativity}

If the energy-momentum tensor of the matter content is ${\mathcal T}_{\mu\nu}$, the New General Relativity \cite{HaSh} equations of motion can be written as \cite{we}
\begin{equation}
\label{NGReq}
 {\mathop\bigtriangledown\limits^{(0)}}_{\alpha} {\mathfrak S}_{\mu}^{\hphantom{\mu}\nu\alpha}- {\mathfrak S}^{\alpha\nu\beta} K_{\alpha\mu\beta} + \frac12 {\mathfrak T}\delta^{\nu}_{\mu}= 8\pi G\cdot {\mathcal T}^{\nu}_{\mu}
\end{equation}
where $K_{\alpha\mu\nu} = - K_{\nu\mu\alpha}$ is the contortion tensor, the difference between the connection (\ref{theconn}) and the Levi-Civita connection of the metric (\ref{themetr}), which can also be written as
$$K_{\alpha\mu\nu}=\frac12 \left(T_{\alpha\mu\nu}+T_{\mu\alpha\nu}+T_{\nu\alpha\mu}\right),$$
and ${\mathop\bigtriangledown\limits^{(0)}}_{\alpha}$ is the operator of Riemannian covariant derivative corresponding to the metric (\ref{themetr}). When $a=b=c=1$, the model reduces to TEGR, and the left-hand side of the equation (\ref{NGReq}) is equal to the usual Einstein tensor $G^{\nu}_{\mu}$. It gets modified by variations of the model parameters, so we will denote it as
\begin{equation}
\label{genEi}
{\mathfrak G}^{\hphantom{\mu}\nu}_{\mu}\equiv  {\mathop\bigtriangledown\limits^{(0)}}_{\alpha} {\mathfrak S}_{\mu}^{\hphantom{\mu}\nu\alpha}- {\mathfrak S}^{\alpha\nu\beta} K_{\alpha\mu\beta} + \frac12 {\mathfrak T}\delta^{\nu}_{\mu}
\end{equation}
with the equations then being ${\mathfrak G}^{\hphantom{\mu}\nu}_{\mu} = 8\pi G\cdot {\mathcal T}^{\nu}_{\mu}$.

It is important to mention that our theories do respect the symmetry under diffeomorphisms. Indeed, the torsion tensor (\ref{torstuff}) for the connection (\ref{theconn}) is defined in terms of antisymmetrised partial derivatives of the differential 1-forms $e^a$. This is a covariant notion, that of differential 2-forms $de^a$. Another way to understand this covariance is via the geometric meaning \cite{meGeom} of the connection (\ref{theconn}). Namely, it ensures that the basis of the vectors $e_a$ is covariantly constant, in terms of this connection. Of course, this symmetry leads to generalised Bianchi identities \cite{HaSh, we, Bianchi}
\begin{equation}
\label{Bianchi}
{\mathop{\bigtriangledown}\limits^{(0)}} {}_{\nu}  {\mathfrak G}^{\mu\nu} + K^{\alpha\mu\beta} {\mathfrak G}_{\alpha\beta}=0.
\end{equation}
which, in particular, show "covariant conservation" of ${\mathfrak G}_{\mu\nu}$ as long as its antisymmetric part is made to vanish.

As is always the case in modified teleparallel models, it is convenient and instructive to separate the equations of motion (\ref{NGReq}) into symmetric and antisymmetric parts. Assuming that the energy-momentum tensor of matter is the usual symmetric tensor, and no hypermomentum is around, the results read
\begin{equation}
\label{asymm}
\frac12 {\mathop{\bigtriangledown}\limits^{(0)}}_{\alpha} \left({\mathfrak S}_{\mu\nu}^{\hphantom{\mu\nu}\alpha}-{\mathfrak S}_{\nu\mu}^{\hphantom{\mu\nu}\alpha}\right) - \frac12 \left( K_{\alpha\mu\beta}{\mathfrak S}^{\alpha\hphantom{\nu}\beta}_{\hphantom{\alpha}\nu} -  K_{\alpha\nu\beta}{\mathfrak S}^{\alpha\hphantom{\mu}\beta}_{\hphantom{\alpha}\mu}\right)=0,
\end{equation}
\begin{equation}
\label{symm}
\frac12 {\mathop{\bigtriangledown}\limits^{(0)}}_{\alpha} \left({\mathfrak S}_{\mu}^{\hphantom{\mu}\nu\alpha}+{\mathfrak S}_{\hphantom{\nu}\mu}^{\nu\hphantom{\mu}\alpha}\right) - \frac12 \left( K_{\alpha\mu\beta}{\mathfrak S}^{\alpha\nu\beta} +  K^{\alpha\nu\beta}{\mathfrak S}_{\alpha\mu\beta}\right) + \frac12 {\mathfrak T} \delta^{\nu}_{\mu}=8\pi G \cdot {\mathcal T}^{\nu}_{\mu}
\end{equation}
of which we will now need only the most elementary first terms. For the sake of brevity, from now on we put $8\pi G=1$. Moreover, we will explicitly treat the case of vacuum only, $ {\mathcal T}^{\nu}_{\mu}=0$.

\subsection{The weak field limit}

As was already explained in the Introduction, the first step towards better understanding the New General Relativity dynamics we are taking now is linearisation of the theory around the  trivial Minkowski background which is 
$$e^a_{\mu} = \delta^a_{\mu}.$$
The tetrad and the co-tetrad are unit matrices, with infinitesimal perturbations around them.

Let us briefly mention that this rather trivial tetrad is by far not the only option for having the Minkowski metric. Every metric possesses a six-dimensional manifold of possible tetrad choices, and those are not equivalent to each other in modified teleparallel theories, unless we go for corresponding transformations of the spin connection, too. In $f(\mathbb T)$ theories those different backgrounds have exhibited different properties of perturbations around \cite{BGKV, nsM}. However, for the much less studied NewGR models, we prefer to start with the simplest choice. Moreover, it is the only choice which satisfies all the Minkowski space symmetries, and has both torsion and Levi-Civita curvature zero.

We denote the perturbations of the co-tetrad as
\begin{equation}
\label{tetpert}
e^a_{\mu}=  I + {\mathfrak e}^a_{\mu}
\end{equation}
with $I$ staying for the unit matrix. In the perturbation variable $\mathfrak e$, we will always keep the position of indices fixed as above (\ref{tetpert}), so that for the tetrad, i.e. the inverse of the co-tetrad, the obvious relation
$$e^{\mu}_a = I - {\mathfrak e}^a_{\mu}+ {\mathcal O}({\mathfrak e}^2)$$
holds. Both types of indices, numbering the 1-forms and marking their components, can take both the time value  (zero) and the spatial values which will be indexed by letters $i,j,k,l$.

All the geometric quantities, except the mere metric itself, are zero at the background level. Therefore,
$${\mathfrak G}^{\hphantom{\mu}\nu}_{\mu}\equiv  {\mathop{\bigtriangledown}\limits^{(0)}}_{\alpha} {\mathfrak S}_{\mu}^{\hphantom{\mu}\nu\alpha} - K_{\alpha\mu\beta}{\mathfrak S}^{\alpha\nu\beta} + \frac12 {\mathfrak T}\delta^{\nu}_{\mu} = \partial_{\alpha} {\mathfrak S}_{\mu}^{\hphantom{\mu}\nu\alpha} + {\mathcal O}(\mathfrak e^2) $$
and the positions of all the indices can be changed by the simple $\eta_{\mu\nu}$. In particular, the generalised Bianchi identities (\ref{Bianchi}) are trivially satisfied at the linear order, 
$$\partial_{\nu}{\mathfrak G}^{\hphantom{\mu}\nu}_{\mu}\equiv 0,$$
 due to antisymmetry of $\mathfrak S_{\mu}^{\hphantom{\mu}\nu\alpha}$ in its last two indices.

For the torsion tensor components (\ref{torstuff}) with lower indices
$$T_{\alpha\mu\nu}=\eta_{ab} e^a_{\alpha} \left(\partial_{\mu} e^b_{\nu} - \partial_{\nu} e^b_{\mu}\right)$$
we conveniently find
\begin{equation}
\label{torpert}
\begin{array}{rcl}
T_{00i}= -T_{0i0} & = &  {\dot{\mathfrak e}}^0_i - \partial_i {\mathfrak e}^0_0,\\
T_{0ij} & = &  \partial_i {\mathfrak e}^0_j - \partial_j {\mathfrak e}^0_i,\\
T_{i0j} =- T_{ij0} & = & - {\dot{\mathfrak e}}^i_j + \partial_j {\mathfrak e}^i_0,\\
T_{ijk} & = & \partial_k {\mathfrak e}^i_j - \partial_j {\mathfrak e}^i_k,
\end{array}
\end{equation}
to the linear order in perturbations. In the same approximation, the metric is given by
\begin{equation}
\label{metpert}
g_{00}=1 + 2 {\mathfrak e}^0_0, \qquad g_{0i}= {\mathfrak e}^0_i - {\mathfrak e}^i_0, \qquad g_{ij}= - \left(\delta_{ij} + {\mathfrak e}^i_j + {\mathfrak e}^j_i\right)
\end{equation}
with its inverse being
$$g^{00}= 1 - 2 {\mathfrak e}^0_0, \qquad g^{0i}= {\mathfrak e}^0_i - {\mathfrak e}^i_0, \qquad g^{ij}= -  \left(\delta_{ij} - {\mathfrak e}^i_j - {\mathfrak e}^j_i\right).$$
And all the teleparallel quantities are simple algebraic combinations of the torsion tensor components (\ref{torpert}). For example, we can find the torsion vector:
\begin{equation}
\label{torvec}
T_0= {\dot{\mathfrak e}}^i_i - \partial_i {\mathfrak e}^i_0, \qquad T_i= - {\dot{\mathfrak e}}^0_i + \partial_i {\mathfrak e}^0_0 - \partial_j {\mathfrak e}^j_i + \partial_i {\mathfrak e}^j_j,
\end{equation}
summation over the repeated spatial indices is assumed. Also for the superpotential components one can find simple formulae such as ${\mathfrak S}_{00i}=\frac{a+b}{2}\cdot T_{00i} + c\cdot T_i$.

Of course, linearised dynamics is always relatively simple. However, we still face quite some number of variables, and it would be very nice to have some order in there. For that, and  having in mind possible future applications, we take the usual approach of cosmological perturbation theory which has also been successfully applied in modified teleparallel models \cite{meTo}. Namely, we separate the perturbations into scalar, vector, and tensor parts (with respect to the spatial rotations). They won't be able to influence each other at the linear level, and therefore can be treated separately. We parametrise the perturbations as follows \cite{meTo}
\begin{equation}
\label{tetrpert}
\begin{array}{rcl}
{\mathfrak e}^0_0 & = & \phi\\
{\mathfrak e}^0_i & = & \partial_i \beta+u_i\\
{\mathfrak e}^i_0 & = & \partial_i \zeta+v_i\\
{\mathfrak e}^i_j & = & -\psi \delta_{ij}+\partial^2_{ij}\sigma+\epsilon_{ijk}\partial_k s+\partial_j c_i+\epsilon_{ijk}w_k+\frac12 h_{ij}.
\end{array}
\end{equation}

Here we have six scalars $\phi,\beta,\zeta,\psi,\sigma,s$, four divergenceless ($\partial_i {\mathbb V}_i =0$) vectors $u_i,v_i,c_i,w_k$, and one tensor mode $h_{ij}$ which is assumed to be symmetric $h_{ij}=h_{ji}$, traceless $h_{ii}=0$, and totally divergenceless $\partial_{i} h_{ij}=0$. Altogether it means $6 + 2 \times 4 +2 = 16$ variables, the total number of the tetrad components. Note that on top of the usual metric perturbations ($\phi$, $\psi$, $\zeta-\beta$, $\sigma$, $u_i-v_i$, $c_i$, $h_{ij}$) of equations  (\ref{metpert}), we have an arbitrary Lorentz transformation there: rotations around $\partial_k s + w_k$ and boosts along $\partial_i (\beta + \zeta) + u_i + v_i$.

\subsection{Gauge symmetry}

As it was already mentioned above, the gauge symmetry at hand, in any of these models, is that of infinitesimal coordinate transformations, $x^{\mu}\longrightarrow x^{\mu} + \xi^{\mu}(x)$.  If the co-tetrad is transformed as a collection of  1-forms $e^a$, then the obvious gauge freedom is
$${e^{\prime}}^a_{\mu}(x^{\prime})=\frac{\partial x^{\nu}}{\partial {x^{\prime}}^{\mu}}\cdot {e}^a_{\nu}(x).$$
Since at the background level, $e^a_{\mu}=\delta^a_{\mu}$, i.e. the tetrad components do  not depend on the coordinates at all, the transformation takes the form of
$${\mathfrak e}^a_{\mu} \longrightarrow {\mathfrak e}^a_{\mu} - \partial_{\mu}\xi^a.$$
In relation to the spatial rotations, the $\xi^0$ is a scalar, while the spatial part can be decomposed into a scalar and a divergenceless vector as $\xi^i=\partial_i\xi + {\tilde \xi}_i$, too. It finally yields
\begin{equation}
\label{gaugetrans}
\begin{array}{rcl}
\phi & \longrightarrow & \phi-{\dot \xi}^0\\
\beta & \longrightarrow & \beta-\xi^0\\
\zeta & \longrightarrow & \zeta-\dot\xi\\
v_i & \longrightarrow & v_i -{\dot {\tilde\xi}_i}\\
\sigma & \longrightarrow & \sigma-\xi\\
c_i & \longrightarrow & c_i -{\tilde\xi}_i.
\end{array}
\end{equation}
Note that it is the usual picture we have in cosmology, with an important simplification in that the Hubble constant $\mathcal H$ terms, ${\mathcal H}\xi^0$, are not here.

The simplest choice of (linearly) gauge invariant variables is
\begin{equation}
\label{gaugeinv}
s, \quad Z=\zeta - \dot\sigma, \quad \Phi=\phi - \dot\beta, \quad \psi, \quad w_i, \quad u_i, \quad V_i=v_i - {\dot c}_i, \quad h_{ij}
\end{equation}
which are $4 + 2\times 3 + 2 = 12$ quantities, precisely as it must be after subtracting $4$ gauge freedoms from $16$ components. As to the metric perturbations, note that (the zero-Hubble-constant limit of) the usual Bardeen variables can be found as
\begin{equation}
\label{Bardeen}
\Phi_B = \Phi + \dot Z, \qquad \Psi_B = \psi, \qquad  {V_B}_i = V_i - u_i,\qquad  h_{ij}.
\end{equation}

Our gauge choice, which is quite usual in its metric part, and was also used in $f(\mathbb T)$ gravity \cite{meTo}, will be
\begin{equation}
\label{ourgauge}
\sigma=0, \qquad \beta=\zeta, \qquad c_i=0
\end{equation}
which is a natural generalisation of the conformal Newtonian gauge of GR. Precisely as the latter, it is very nice because it fully fixes the gauge and its remaining non-invariant quantities $\phi, \zeta, v_i$ are numerically equal (in this particular gauge) to the gauge-invariant variables, $\Phi_B, Z, V_i$ respectively.

\section{The tensor sector}

We start from the tensor sector of perturbations because it is where the usual GWs of GR reside, and as always, it is also the simplest one. In particular, the only non-vanishing torsion tensor components (\ref{torpert}) are
\begin{equation}
\label{tentor}
\begin{array}{rcl}
T_{ijk} & =  & \frac12 \left(\partial_k h_{ij} - \partial_j h_{ik}\right),\\
T_{i0j} & = &  - \frac12 {\dot h}_{ij} \vphantom{\left(\dot H\right)},
\end{array}
\end{equation}
with no contribution to the torsion vector, nor any other vector or scalar quantity. It is gauge invariant and can contribute to the symmetric spatial equation (\ref{symm}) only, 
$$\frac12 \left({\mathfrak G}_i^{\hphantom{i}j}(h)+ {\mathfrak G}^j_{\hphantom{j}i}(h)\right) =\mathop{{{\mathcal  T}_i^j}}\limits_{\mathrm{(tensor)}}.$$

Substituting the torsion tensor components (\ref{tentor}), we find
\begin{equation}
\label{GWmat}
{\mathfrak G}_i^{\hphantom{i}j} = - \frac{ a+b}{4}\cdot \left({\ddot h}_{ij} -\bigtriangleup h_{ij}\right)= \mathop{{{\mathcal  T}_i^j}}\limits_{\mathrm{(tensor)}}.
\end{equation}
Therefore, in absence of matter, the GWs' equation of motion is
\begin{equation}
\label{GWeq}
( a+b)\cdot \left({\ddot h}_{ij} -\bigtriangleup h_{ij}\right)= 0.
\end{equation}
In other words, the spin two modes are the usual ones, with the standard wave equation. Since we consider the model in vacuum, it has no source term. Though, precisely in the case of tensor modes, this fact is more general. In a spatially isotropic and homogeneous cosmology, the gravitational waves have no source term either, as long as the matter content can be approximated as an ideal fluid.

Note also that the equation (\ref{GWmat}) for these waves is just the same as in GR, with the only change in renormalising the strength of their coupling to matter due to the $\frac{a+b}{2}$ factor as compared to the standard gravity. There is a case though, $a+b=0$, when the linear spin-2 waves do not exist at all, representing yet another gauge freedom. The reason is of course very simple. The Lagrangian of such a model depends on the torsion vector and axial vector only, with no tensorial part. In this respect, the paper \cite{GWtel} is inaccurate when stating that the "two tensorial modes of general relativity are always present".

\section{The scalar sector}

Now we turn to another cosmologically important sector, the scalar one. In our gauge it turns out to be relatively simple, too. The torsion components (\ref{torpert}) are
\begin{equation}
\label{scaltor}
\begin{array}{rcl}
T_{00i} & = & \partial_i \left({\dot\beta} - \phi \right)\\
{} & {} & \xrightarrow{\text{gauge fixing}}  \partial_i \left({\dot\zeta} - \phi\right)\\
T_{ijk} & = &  \delta_{ik}\partial_j \psi - \delta_{ij}\partial_k \psi + \epsilon_{ijl} \partial^2_{kl} s - \epsilon_{ikl} \partial^2_{jl} s \\
T_{i0j} & = & {\dot\psi} \delta_{ij} - \partial^2_{ij} {\dot\sigma} + \partial^2_{ij} \zeta  - \epsilon_{ijk}\partial_k {\dot s}  \\
{} & {} & \xrightarrow{\text{gauge fixing}} {\dot\psi} \delta_{ij} + \partial^2_{ij} \zeta - \epsilon_{ijk}\partial_k {\dot s} 
\end{array}
\end{equation}
for the tensor, and
\begin{equation}
\label{scalvec}
\begin{array}{rcl}
T_0 & = &  \bigtriangleup \left({\dot\sigma} - \zeta\right) -3 \dot\psi\\
{} & {} & \xrightarrow{\text{gauge fixing}}  - \bigtriangleup\zeta -3 \dot\psi\\
T_i & = &  \partial_i \left(\phi - {\dot\beta} - 2\psi\right)\\
{} & {} & \xrightarrow{\text{gauge fixing}} \partial_i \left(\phi - {\dot\zeta} - 2\psi\right) 
\end{array}
\end{equation}
for the vector. Note that, in the initial expressions, the variables $\zeta$ and $\phi$ had no velocities. This is the scalar part of the diffeomorphism invariance producing one half of the necessary primary constraints which do not depend on the particular parameters of the model at hand.

The antisymmetric part (\ref{asymm}) of equations is
\begin{equation}
\label{asymscal}
\begin{array}{rcccl}
\frac{{\mathfrak G}_{0i}-{\mathfrak G}_{i0}}{2} & = & \frac{2c- a-b}{4}\cdot \partial_i\left({\ddot\zeta}-\bigtriangleup\zeta - {\dot\phi} -{\dot\psi}\right) & = & 0\\
\frac{{\mathfrak G}_{ij}-{\mathfrak G}_{ji}}{2} & = & \frac{ a-b}{2}\cdot \epsilon_{ijk}\partial_k ({\ddot s} - \bigtriangleup s) & = & 0,
\end{array}
\end{equation}
with the symmetric part (\ref{symm}) taking the form of
\begin{equation}
\label{symscal}
\begin{array}{rcl}
\bigtriangleup \left(2c\cdot \psi + \frac{2c-a-b}{2}\cdot ({\dot\zeta}-\phi)\right) & = &\mathop{{{\mathcal  T}_0^0}}\limits_{\mathrm{(scalar)}}\\
\partial_i \left(\frac{10c- a-b}{4}\cdot {\dot\psi} + \frac{2c-a-b}{4}\cdot ({\ddot\zeta} + \bigtriangleup\zeta - {\dot\phi})\right) & = &  \mathop{{{\mathcal  T}_i^0}}\limits_{\mathrm{(scalar)}}\\
 \partial^2_{ij}\left(c\cdot \phi -\frac{4c -a-b}{2}\cdot \psi -\frac{2c-a-b}{2}\cdot {\dot\zeta}\right) - \delta_{ij}\left(\frac{6c -a-b}{2}\cdot {\ddot\psi} - \frac{4c - a-b}{2}\cdot \bigtriangleup\psi + c\cdot \bigtriangleup\phi \right) & = & \mathop{{{\mathcal  T}_i^j}}\limits_{\mathrm{(scalar)}}
\end{array}
\end{equation}
for the temporal, mixed, and spatial components respectively.

Obviously, in the GR case of $a=b=c$, the antisymmetric part (\ref{asymscal}) of equations disappears, as well as any trace of Lorentz perturbations in the symmetric part (\ref{symscal}). Also, putting $a=b=c=1$ and restoring the gravitational constant, one can see that these equations coincide with the Minkowski limit of the standard cosmological perturbations, in particular with the standard gravity case of $\bigtriangleup\psi = 4\pi G \rho$.

As always in perturbation theory, if a spatial gradient of a scalar is equal to zero, we solve it as the scalar being just zero. A scalar of zero Laplacian is also taken as being zero. In other words, we consider only local perturbations which are everywhere small. In particular, the spatial components of symmetric equations (\ref{symscal}) are treated as two independent equations, the $\delta_{ij}$ and $\partial^2_{ij}$ parts separately. Therefore, we get the following system of equations in vacuum:
\begin{equation}
\label{scaleq}
\begin{array}{lrcl}
\mathrm a) & (2c - a- b)\cdot \left({\ddot\zeta}-\bigtriangleup\zeta - {\dot\phi} -{\dot\psi}\right)  & = & 0\\
\mathrm b) & (a - b)\cdot  \left({\ddot s} - \bigtriangleup s \vphantom{\dot\zeta}\right) & = & 0\\
\mathrm c) & 4c\cdot \psi + (2c-a-b)\cdot \left({\dot\zeta}-\phi\right)  & = & 0\\
\mathrm d) \hphantom{some space} & (10c- a-b)\cdot {\dot\psi} + (2c-a-b)\cdot \left({\ddot\zeta} + \bigtriangleup\zeta - {\dot\phi}\right)  & = & 0\\
\mathrm e) & 2c\cdot \phi - (4c -a-b)\cdot \psi - (2c-a-b) \cdot {\dot\zeta} \vphantom{\left(\dot\zeta\right)}  & = & 0\\
\mathrm f) & (6c -a-b)\cdot {\ddot\psi} - (4c - a-b)\cdot \bigtriangleup\psi + 2c\cdot \bigtriangleup\phi  \vphantom{\left(\dot\zeta\right)} & = & 0.
\end{array}
\end{equation}

Note in passing that a constant $\partial \xi$, or a constant $\partial s$, does not contribute to the equations at all, from the very beginning. This is because a global Lorentz transformation is always a symmetry of teleparallel actions. We ignore such options, in all the variables. A possible spatially homogeneous mode in the Newtonian potentials for cosmology can be treated as an improper choice of the background. Even though the meaning of constant Lorentz rotations might probably be more puzzling, we will neglect them, too. In itself, it might be a topic of separate investigation, for it can have a direct connection to strange modes with reduced Cauchy data in $f(\mathbb T)$ models around non-standard tetrads with Minkowski metric \cite{nsM}.

From the equation (\ref{scaleq}b) we immediately see that the pseudoscalar mode
\begin{equation}
\label{pseq}
 (a - b)\cdot  \left({\ddot s} - \bigtriangleup s \vphantom{\dot\zeta}\right) = 0,
\end{equation}
 which describes spatial rotations of the tetrad, obeys the standard wave  equation, except for the case of $a=b$ when it gets eaten by a gauge freedom, and does not appear anywhere else. This is yet another "polarisation" of gravitational waves which is in a fully stealth mode since it does not influence the metric, and therefore is undetectable at the level of linear perturbations and test bodies of the usual type.

One might then worry that the five remaining equations for the three variables ($\phi,\psi,\zeta$) could have only the zero solution. Actually, it would be indeed the case for scalar modes in vacuum GR. However, it is not the case now due to the new dynamics in New GR models. The system (\ref{scaleq}) is not overdetermined because of the Bianchi identity $\partial_{\nu} {\mathfrak G}^{\hphantom{\mu}\nu}_{\mu}=\partial^2_{\alpha\beta} {\mathfrak S}_{\mu}^{\hphantom{\mu}\alpha\beta}\equiv 0$. In particular, the temporal component of the identity presents itself as the time derivative of (\ref{scaleq}c) being equal to one half the sum of (\ref{scaleq}a) and (\ref{scaleq}d). Then, out of the latter two equations, it suffices to keep only (half) their difference, or roughly the ${\mathfrak G}_{i0}$ component:
\begin{equation}
\label{scali0}
(2c-a-b)\cdot \bigtriangleup\zeta + (6c -a -b)\cdot \dot\psi =0.
\end{equation}
After that, thanks to the spatial components of the Bianchi identity, we can see that the time derivative of this equation (\ref{scali0}) reproduces the equation (\ref{scaleq}f) minus a Laplacian of (\ref{scaleq}e).

Therefore, for finding the dynamical perturbations, we can safely restrict ourselves to only three equations for the $\phi,\psi,\zeta$ variables,  (\ref{scaleq}c), (\ref{scali0}), and (\ref{scaleq}e):
\begin{equation}
\label{finscal}
\begin{array}{rcl}
4c\cdot \psi + (2c-a-b)\cdot \left({\dot\zeta}-\phi\right)  & = & 0\\
(2c-a-b)\cdot \bigtriangleup\zeta + (6c -a -b)\cdot \dot\psi \vphantom{\left(\dot\zeta\right)}   & = & 0\\
2c\cdot \phi - (4c -a-b)\cdot \psi - (2c-a-b) \cdot {\dot\zeta} \vphantom{\left(\dot\zeta\right)}  & = & 0.
\end{array}
\end{equation}
 Note that normally, as we have also done it here, one should better neglect the higher time derivative options when removing dependent equations, in order to not lose a constraint.

When dealing with the system of equations (\ref{finscal}), let us first consider the general situation of
$$a+b \neq 0,\qquad 2c - a -b \neq 0,\qquad 6c -a -b \neq 0.$$
The sum of the first and the third equations gives us
$$4(a+b)\cdot (\phi+\psi)=0$$
which implies $\phi=-\psi$. Then the equations  (\ref{finscal}) produce
\begin{equation*}
\begin{array}{rcl}
 (2c-a-b)\cdot {\dot\zeta} & = &  (6c -a -b)\cdot \phi \vphantom{\frac{A}{\int}} \\
(2c-a-b)\cdot \bigtriangleup\zeta & = &  (6c -a -b)\cdot \dot\phi \vphantom{\frac{\int}{B}}.
\end{array}
\end{equation*}
These relations tell us that the variables must obey the standard wave equation, and we have reduced the system (\ref{finscal}) to
$${\ddot\zeta} - \bigtriangleup\zeta =0, \qquad \phi=-\psi=\frac{2c-a-b}{6c-a-b}\cdot \dot\zeta.$$
In other words, there is one new dynamical mode related to the Lorentz boosts $\zeta$ which is observable as a conformal ($\phi=-\psi$) mode in the metric.

Let us now consider the special cases. If to be thinking of a model close to GR, the most important special option is $2c=a+b$. In this case, on top of the conformal character $\phi=-\psi$, we get the standard $\phi=\psi$ requirement of GR from the third equation of the system (\ref{finscal}). Altogether, $\zeta$ is then arbitrary, i.e. yet another gauge freedom, and $\phi=\psi=0$ with no new mode in this sector, that is precisely the same as in GR. Of course, for that we have assumed that $2c=a+b\neq0$, for otherwise all these variables are free.

Coming to unrealistic cases, if $6c=a+b\neq0$ in the equations (\ref{finscal}), then $\zeta=0$ while $\phi=-\psi$ but otherwise arbitrary which means that there is no dynamical mode, however the conformal mode of the metric enjoys a gauge freedom, so that the behaviour of massive test particles is unpredictable. Finally, if $a+b=0$ though $c\neq 0$, then the metric is again not fully predictable. In the set of equalities (\ref{finscal}), there appear to be only two independent equations, $3\dot\psi = - \bigtriangleup\zeta$ and $\dot\zeta = \phi - 2\psi$. We can take it as $\psi$ being gauge free, while $\phi$ and $\zeta$ fully constrained by $ \bigtriangleup\phi=-3{\ddot\psi} + 2 \bigtriangleup\psi$ and $\bigtriangleup\zeta=-3\dot\psi$.

\section{The vector sector}

Finally, in the vector sector, the torsion tensor components (\ref{torpert}) take the form of
\begin{equation}
\label{vecttor}
\begin{array}{rcl}
T_{00i} & = & {\dot u}_i,\\
T_{0ij} & = &  \partial_i u_j - \partial_j u_i,\\
T_{ijk} & = &  \epsilon_{ijl}\partial_k w_l - \epsilon_{ikl}\partial_j w_l,\\
T_{i0j} & = &  \partial_j v_i  - \partial_j {\dot c}_i - \epsilon_{ijk}{\dot w}_k, \\
{} & {} & \xrightarrow{\text{gauge fixing}} \partial_j v_i - \epsilon_{ijk}{\dot w}_k
\end{array}
\end{equation}
with the torsion vector equal to
\begin{equation}
\label{vectvec}
 T_i = - {\dot u}_i + \epsilon_{ijk} \partial_j w_k.
\end{equation}
Analogously to the case of scalars,  there is a vector $v_i $ which has no time derivatives in the geometric quantities, thus producing another half of the guaranteed primary constraints from diffeomorphisms.

For the antisymmetric equation (\ref{asymm}) we get
\begin{equation}
\label{asymvect}
\begin{array}{rcl}
\frac{2c- a-b}{4}\cdot {\ddot u}_i + \frac{a -b}{4}\cdot \bigtriangleup (u_i + v_i) -\frac{2c+a-3b}{4}\cdot \epsilon_{ijk}\partial_j {\dot w}_k & = & 0 \vphantom{\frac{A}{\int}}\\
\frac{a -b}{2}\cdot \epsilon_{ijk} ({\ddot w}_k- \bigtriangleup w_k) +\frac{c-b}{2}\cdot (\partial_i {\dot u}_j-\partial_j {\dot u}_i) + \frac{a -b}{4} \cdot (\partial_i {\dot v}_j-\partial_j {\dot v}_i) + \frac{2c+a-3b}{4}\cdot (\epsilon_{ikl}\partial^2_{jk} w_l - \epsilon_{jkl}\partial^2_{ik} w_l) & = & 0 \vphantom{\frac{\int}{B}}
\end{array}
\end{equation}
in the mixed components $\frac12 ({\mathfrak G}_{0i}-{\mathfrak G}_{i0})$ and in the spatial parts $\frac12 ({\mathfrak G}_{ij}-{\mathfrak G}_{ji})$ respectively. The symmetric equation  (\ref{symm}) acquires a nice form of
\begin{equation}
\label{symvect}
\begin{array}{rcl}
\frac{ a+b}{4}\cdot \bigtriangleup (u_i - v_i) + \frac{2c-a-b}{4}\cdot \left({\ddot u}_i - \epsilon_{ijk} \partial_j {\dot w}_k\right) & = &  \mathop{{{\mathcal  T}_i^0}}\limits_{\mathrm{(vector)}}\\
\frac{a+b}{4}\cdot (\partial_i {\dot v}_j + \partial_j {\dot v}_i) - \frac{c}{2}\cdot (\partial_i {\dot u}_j + \partial_j {\dot u}_i) + \frac{2c- a-b}{4}\cdot (\epsilon_{ikl}\partial^2_{jk} w_l + \epsilon_{jkl}\partial^2_{ik} w_l) & = & \mathop{{{\mathcal  T}_i^j}}\limits_{\mathrm{(vector)}}.
\end{array}
\end{equation}
The case of GR is also easily recognisable when $a=b=c$.

In order to facilitate the discussion, let's parametrise the divergenceless vector $w_i$ as a curl of another:
$$w_i=\epsilon_{ijk} \partial_j \chi_k, \qquad \partial_i \chi_i =0.$$
After that, the spatial components of equations (\ref{asymvect}, \ref{symvect}) acquire the form of $\partial_i {\mathbb V}_j \pm \partial_j {\mathbb V}_i =0$. By taking a divergence, such equations imply $\bigtriangleup {\mathbb V}_i =0$ which, as always, we solve as ${\mathbb V}_i =0$. Altogether it allows us to rewrite the equations (\ref{asymvect}) and (\ref{symvect}) in vacuum as follows:
\begin{equation}
\label{vecteq}
\begin{array}{lrcl}
\mathrm a) & (2c- a-b)\cdot {\ddot u}_i + (a -b)\cdot \bigtriangleup (u_i + v_i) + (2c+a-3b)\cdot \bigtriangleup {\dot\chi}_i \vphantom{\left(\dot\zeta\right)} & = & 0\\
\mathrm b) & 2(a - b)\cdot {\ddot \chi}_i + (2c-a-b)\cdot \bigtriangleup \chi_i + 2(c-b)\cdot {\dot u}_i + (a-b)\cdot {\dot v}_i \vphantom{\left(\dot\zeta\right)}& = & 0\\
\mathrm c) & (2c- a-b)\cdot {\ddot u}_i + (a+b)\cdot  \bigtriangleup (u_i - v_i) + (2c-a-b)\cdot \bigtriangleup {\dot\chi}_i & \vphantom{\left(\dot\zeta\right)} = & 0\\
\mathrm d) \hphantom{some big space} & (a+b)\cdot {\dot v}_i -2c\cdot  {\dot u}_i - (2c-a-b)\cdot \bigtriangleup \chi_i  & \vphantom{\left(\dot\zeta\right)} = & 0.
\end{array}
\end{equation}

Precisely as in the scalar sector, due to having fixed the gauge, there are less variables than equations. However, this is taken care of by Bianchi identities (\ref{Bianchi}) again. If we subtract (\ref{vecteq}c) from (\ref{vecteq}a), the result
\begin{equation}
\label{vecti0}
2\bigtriangleup \left(a\cdot v_i - b\cdot  u_i +(a-b)\cdot {\dot\chi}_i \vphantom{\int}\right)=0
\end{equation}
is basically the ${\mathfrak G}_{i0}$ component. And if we take a time derivative of this equation (\ref{vecti0}), this is precisely what one gets by taking a Laplacian of the sum of (\ref{vecteq}b) and (\ref{vecteq}d).

Then we can safely neglect the equation (\ref{vecteq}b) and substitute the mixed components equations, that is  (\ref{vecteq}a) and (\ref{vecteq}c), by their sum and difference (\ref{vecti0}), i.e. ${\mathfrak G}_{0i}$ and ${\mathfrak G}_{i0}$ instead of symmetrised and antisymmetrised versions. Therefore, our system of equations reads
\begin{equation}
\label{finvect}
\begin{array}{rcl}
(2c- a-b)\cdot {\ddot u}_i +a\cdot \bigtriangleup u_i - b\cdot \bigtriangleup v_i+ 2(c-b)\cdot \bigtriangleup {\dot\chi}_i \vphantom{\left(\dot\zeta\right)}   & = & 0\\
a\cdot v_i - b\cdot  u_i +(a-b)\cdot {\dot\chi}_i \vphantom{\left(\dot\zeta\right)}   & = & 0\\
(a+b)\cdot {\dot v}_i -2c\cdot {\dot u}_i - (2c-a-b)\cdot \bigtriangleup \chi_i \vphantom{\left(\dot\zeta\right)}  & = & 0.
\end{array}
\end{equation}
If we combine the last two equations for finding $2(c-b)\cdot \bigtriangleup {\dot\chi}_i $ and then substitute the result into the first one, then the standard wave equation for the metric perturbation emerges, $\square (u_i - v_i)=0$. It gives us a hint that it might be beneficial to organise the mixed tetrad components into the metric and the Lorentz boost parts
$${\mathcal M}_i = \frac{u_i -v_i}{2}, \qquad {\mathcal L}_i = \frac{u_i +v_i}{2}$$
which brings the equations (\ref{finvect}) to their final shape
\begin{equation}
\label{superfin}
\begin{array}{rcl}
(a+b)\cdot \left({\ddot{\mathcal M}}_i- \bigtriangleup{\mathcal M}_i\right) & = & 0\\
(a+b)\cdot {\mathcal M}_i - (a-b)\cdot \left({\mathcal L}_i + {\dot\chi}_i  \vphantom{{\ddot{\mathcal M}}_i}\right) & = & 0\\
(a+b+2c)\cdot {\dot{\mathcal M}} - (a+b-2c)\cdot \left({\dot{\mathcal L}}_i + \bigtriangleup \chi_i\right) & = & 0.
\end{array}
\end{equation}

We start again from the generic case of
$$a+b \neq 0, \qquad a-b \neq 0, \qquad 2c-a-b\neq 0.$$
The first equation of the system (\ref{superfin}) describes the metric dynamical mode. At the same time, excluding the ${\mathcal L}_i$ from the last two equations, we get another dynamical equation
\begin{equation}
\label{chieq}
{\ddot\chi}_i - \bigtriangleup \chi_i = \frac{2b(a+b)-4ac}{(a-b)(a+b-2c)}\cdot {\dot{\mathcal M}}_i
\end{equation}
with a source term. Thus we have got two dynamical divergenceless vectors, i.e. four dynamical degrees of freedom, and a constraint for 
$${\mathcal L}_i=\frac{a+b}{a-b}\cdot {\mathcal M}_i - {\dot\chi}_i.$$ 
One can easily check that this is the general solution of the dynamical system (\ref{superfin}). The first equation requires the free wave behaviour of ${\mathcal M}_i$, then the second one gives the constraint for ${\mathcal L}_i$ which, upon substitution into the third equation, shows that everything is solved if and only if the equation (\ref{chieq}) is satisfied.

For the special cases, if we take $a-b=a+b-2c=0$, while keeping $a+b\neq 0$, it is the case of GR with the metric fixed ${\mathcal M}_i=0$ and the Lorentzian variables $\chi_i, {\mathcal L}_i$ gauge free. The case of $a=b\neq 0$ but $2c\neq a+b$ still requires ${\mathcal M}_i=0$ but the Lorentzian sector is only half free with a constraint $\bigtriangleup\chi_i=-{\dot{\mathcal L}}_i$. In the opposite situation of $2c=a+b$ but $a\neq\pm b$, we have a similar story, that is ${\mathcal M}_i=0$ (recall that we always solve $\bigtriangleup{\mathcal M}_i=0$ as ${\mathcal M}_i=0 $ only) and half freedom in the Lorentzian sector restricted by another constraint, ${\mathcal L}_i=-{\dot\chi}_i$.

Finally, the special unrealistic realm is about $a+b=0$. When not imposing any other restriction, it works precisely as in the generic case relating the vectors as  (\ref{chieq}), $ {\dot{\mathcal M}}_i = {\ddot\chi}_i - \bigtriangleup \chi_i$, and  ${\mathcal L}_i=-{\dot\chi}_i$. Since there is no more information, we can treat $\chi_i$ as gauge free while ${\mathcal L}_i$ and ${\mathcal M}_i$ as constrained and half-constrained respectively. If we specify it even more, $a=b=0$ but $c\neq 0$, then there are two vectorial gauge freedoms, including the unpredictability of the metric and a constraint of $\bigtriangleup\chi_i=-{\dot{\mathcal M}}_i-{\dot{\mathcal L}}_i$. If to the contrary, we go for non-zero $a$ and $b$ but $a+b=c=0$, then similarly all is free except for a constraint ${\mathcal L}_i=-{\dot\chi}_i$.

\section{On the degrees of freedom}

Actually, one can find the kinetic matrix of the theory by leaving only the velocities in the torsion tensor (\ref{torpert}) components and calculating the scalar $\mathfrak T$ then (\ref{torsionscalar}). The result in our gauge (\ref{ourgauge}) is
\begin{equation}
\label{kinmat}
{\mathfrak K}=\frac{a+b}{8}\cdot {\dot h}^2_{ij}+ \frac{2c-a-b}{2}\cdot \left({\dot u}^2_i+(\partial_i \dot\zeta)^2\right) +(a-b) \cdot \left({\dot w}^2_i +(\partial_i {\dot s} \vphantom{\dot{\zeta}})^2\right) - \frac{3(6c-a-b)}{2}\cdot {\dot\psi}^2.
\end{equation}
Out of 12 variables in our gauge, 9 have their appearance in the kinetic matrix $\mathfrak K$, two spin-two polarisations with the prefactor of $a+b$, three rotational modes with the prefactor of $a-b$, three boost modes with the prefactor of $2c-a-b$, and one conformal mode with the prefactor of $a+b-6c$. The zeros of these prefactors correspond to various special cases of the model.

In the case of general relativity, $a=b=c=1$, the kinetic matrix (\ref{kinmat}) takes the form of
$${\mathfrak K}_{\mathrm{GR}}=\frac14  {\dot h}^2_{ij} - 6{\dot\psi}^2$$
which is the energy of gravitational waves plus the negative kinetic energy of the conformal mode. The latter isn't a problem for GR because the mode is constrained and not dynamical, however it generically revives as the Boulware-Deser ghost when inaccurately giving a mass to the graviton \cite{Claudia, memass}. Note that in our current case, unlike in massive gravity, since the diffeomorphisms aren't broken, we can hope that the problematic mode is still non-dynamical with no extra effort required for that, while all the rest can be made positive by assuming that $a>|b|$ and $c>\frac{a+b}{2}$. 

\subsection{Comparison with the works on primary constraints}

Various terms in the kinetic matrix (\ref{kinmat}) vanishing correspond to primary constraints in previous works \cite{Dan1, Shy}, up to the change of notations. Their coefficients $c_i$ are $c_1=\frac{a}{4}$, $c_2=\frac{b}{2}$, $c_3=-c$ in our terms. The only difference in this respect is that generically they have 12 variables with time derivatives \cite{Dan1, Shy} in the action, as opposed to our 9, and even more importantly, they have five momenta \cite{Shy} vanishing at $a+b=0$, or their $2c_1+c_2=0$, instead of our two transverse traceless modes. This is because we have fixed the gauge. First, we put $\beta=\zeta$ which is not important at all for now. Out of these two variables only $\beta $ had a time derivative in the action, and we simply renamed it to $\dot\zeta$. The three velocities which disappeared (\ref{ourgauge}) are of $\sigma$ and $c_i$. We can easily check that they are precisely the ones responsible for the three missing $a+b$ prefactors. Indeed, with some integrations by parts, we immediately see that, if not to fix any gauge, the kinetic matrix gets
\begin{multline*}
- \frac{3(6c-a-b)}{2}\cdot {\dot\psi}^2 + (6c-a-b)\cdot {\dot\psi} \bigtriangleup {\dot\sigma} + \frac{a+b-2c}{2}\cdot (\bigtriangleup\dot\sigma)^2\\
 = - \frac{3(6c-a-b)}{2}\cdot \left(\dot\psi - \frac13 \bigtriangleup\dot\sigma\right)^2 +\frac{a+b}{3}\cdot (\bigtriangleup\dot\sigma)^2
\end{multline*}
instead of $ - \frac{3(6c-a-b)}{2}\cdot {\dot\psi}^2$, and
\begin{multline*}
\frac{a}{2}\cdot (\partial_i {\dot c}_j)^2 - (a-b)\cdot \epsilon_{ijk} (\partial_i {\dot c}_j) {\dot w}_k + (a-b)\cdot {\dot w}^2_i  = \frac{a}{2}\cdot (\partial_i {\dot c}_j)^2 - (a-b)\cdot  (\partial_i {\dot c}_j)  (\partial_i {\dot \chi}_j) + (a-b)\cdot (\partial_i {\dot\chi}_j)^2\\
=(a-b)\cdot \left(\partial_i\left({\dot\chi}_j-\frac12 {\dot c}_j\right)\right)^2 + \frac{a+b}{4} \cdot (\partial_i {\dot c}_j)^2
\end{multline*}
instead of $ (a-b)\cdot {\dot w}^2_i $. Therefore, in our analysis we clearly see the presence of all the same primary constraints as in the previous works \cite{Dan1, Shy}.

\subsection{The nine different types of models}

Being fully free in the choice of parameters, the easiest thing to analyse is the case of $a=b=c=0$. This is an empty model of an identically zero action with no dynamical modes, no Lagrangian constraints, and all the variables enjoying the total gauge freedom. The canonical Hamiltonian can be taken equal to zero, and all the constraints are primary first-class ones telling us that all canonical momenta vanish. This is of no interest, of course. Let us now present the nine classes of non-trivial models, restricting ourselves to the linearised dynamics around the trivial Minkowski solution only.

We remind the reader that the standard cosmological tetrad decomposition (\ref{tetrpert}) is used, with the vectorial part of spatial rotations represented as $w_i=\epsilon_{ijk} \partial_j \chi_k$ and new combinations ${\mathcal M}_i = \frac12 (u_i -v_i)$, ${\mathcal L}_i = \frac12 (u_i +v_i)$ of the spatial-temporal variables  introduced. We display the equations below in, let's say, a semi-solved shape, so that the dynamical structure is clear. Note that, after we have fixed the gauge (\ref{ourgauge}), there are 6 metric and 6 Lorentzian variables:
$$\phi, \quad \psi, \quad {\mathcal M}_i, \quad h_{ij}; \qquad \zeta,\quad s,\quad {\mathcal L}_i,\quad \chi_i.$$
Those are the explicit variables for all the models we discuss. And in every case, one can add 4 diffeo gauge symmetries to the lists below.

Note also that the structure of different models depends on restrictions imposed on the model parameters. The four important restrictions are
\begin{equation}
\label{constr}
{\mathrm I})\quad 2c-a-b=0, \quad \mathrm{II})\quad a-b=0, \quad \mathrm{III})\quad a+b=0, \quad \mathrm{IV})\quad 6c-a-b=0.
\end{equation}
For the tensors (\ref{GWeq}), only the restriction III plays a role, for the pseudoscalar (\ref{pseq}) -- only the restriction II, for the rest of the scalars (\ref{finscal}) -- all the restrictions except II, for the vectors (\ref{superfin}) -- all the restrictions except IV. All possible non-trivial options boil down to the cases below.

\begin{center}
{\bf The number of modes after having fixed the diffeo gauge}

\begin{tabular}{c|c|c|c|c|} 
 \hline
& the model according to (\ref{constr}) & dynamical modes & constrained modes &  pure gauge modes\\ 
\hline
6.2.1 & Generic & 8 & 4 & 0\\ 
\hline
6.2.2 & Only I & 3 & 6 & 3\\
\hline
6.2.3 & Only II & 3 & 6 & 3\\
\hline
6.2.4 & Only III & 2 & 5 & 5\\
\hline
6.2.5 & Only IV & 7 & 4 & 1\\
\hline
6.2.6 & Only I and II & 2 & 4 & 6\\
\hline
6.2.7 & Only II and IV & 2 & 6 & 4\\
\hline
6.2.8 & Only II and III & 0 & 4 & 8\\
\hline
6.2.9 & Only I and III and IV & 1 & 2 & 9\\
\hline
\end{tabular}
\end{center}

\subsubsection{The $|a|\neq|b|$, $2c\neq a+b$, $6c\neq a+b$ model}

We start with the most general model, with no restrictions (\ref{constr}) satisfied. All the other options are lower dimensional surfaces in the space of parameters. Let us collect here the full set of equations in the most transparent form:
\begin{equation*}
\left\{
\begin{array}{l}
{\ddot\zeta} - \bigtriangleup\zeta =0,\\
\phi=-\psi=\frac{2c-a-b}{6c-a-b}\cdot \dot\zeta,\\
{\ddot s} - \bigtriangleup s =0,\\
{\ddot{\mathcal M}}_i- \bigtriangleup{\mathcal M}_i =0,\\
{\ddot\chi}_i - \bigtriangleup \chi_i = \frac{2b(a+b)-4ac}{(a-b)(a+b-2c)}\cdot {\dot{\mathcal M}}_i,\\
{\mathcal L}_i=\frac{a+b}{a-b}\cdot {\mathcal M}_i - {\dot\chi}_i,\\
{\ddot h}_{ij} -\bigtriangleup h_{ij}=0.
\end{array}
\right.
\end{equation*}
We see that there is no additional gauge freedom on top of the obligatory diffeomorphisms which we have already gauge-fixed. One can take $\zeta, s, {\mathcal M}_i, \chi_i, h_{ij}$ as dynamical modes, with $\phi, \psi, {\mathcal L}_i$ being constrained. Therefore, out of 12 physical variables, 8 are dynamical and 4 are constrained. The latter are precisely the secondary constraints coming from the four diffeomorphism gauge freedoms "hitting twice" \cite{meHam}.

In our opinion, this model is actually very promising. It seems quite plausible that the diffeomorphisms do "hit twice" at the fully non-linear level, too, and therefore the linear analysis has most probably revealed all the dynamical modes, with no strong coupling behind. As to the potential problem of ghosts, if we impose the restrictions of $a>|b|$ and $c>\frac{a+b}{2}$ on the model parameters, all kinetic energies (\ref{kinmat}) are bounded from below, except for the constrained conformal variable $\psi$. Moreover, if the parameters are not too far from the GR values, so that $2c-a-b$ is much less than $6c-a-b$, then according to the constraint determining $\psi$ in terms of $\dot\zeta \sim \partial\zeta$, the amount of the kinetic energy of $\partial\zeta$ wins over that of $\psi$, therefore keeping the whole quantity (\ref{kinmat}) positive when averaged over time.

This opinion of ours goes against the common belief that a viable model would require $2c=a+b$. It starts from the very first paper \cite{HaSh} which preferred having the same static spherically symmetric solutions as in GR. Since then, however, many works have been done precisely for changing that in order to mimick the effects of Dark Matter by modifying gravity. Another influential objection to these models with $2c\neq a+b$ came in the book \cite{Ortin}. It notices that the new fields influence the metric modes (in the language of Fierz-Pauli and Kalb-Ramond there) and claims it to be phenomenologically unviable. In a sense, it goes in the same direction as objections of the paper \cite{HaSh}. And even more generally, we would say that it is nowhere close to a reasonable motivation for immediately rejecting a modified gravity model. When we have a model with new degrees of freedom, it is nothing but natural that they should interact with the old ones.

\subsubsection{The $2c=a+b$, $|a|\neq|b|$ model}

Now we consider codimension-one submanifolds in the parameter space, and start from the options which can have the values of model parameters not very far  from GR . Our first choice, the restriction I of $2c=a+b$, is the one which has the same static spherically symmetric solutions as in GR \cite{HaSh, we, Obukhov}. The equations are
\begin{equation*}
\left\{
\begin{array}{l}
\phi=\psi=0,\\
{\ddot s} - \bigtriangleup s =0,\\
\mathcal M_i =0,\\
{\mathcal L}_i= - {\dot\chi}_i,\\
{\ddot h}_{ij} -\bigtriangleup h_{ij}=0.
\end{array}
\right.
\end{equation*}
The metric variables behave like in GR, two dynamical modes and four constrained (to zero) variables. The Lorentzian sector has one dynamical mode $s$ and one gauge freedom $\zeta$ in the scalars, and on top of that two more gauge freedoms and two constraints which can be taken as $\chi_i$ and ${\mathcal L}_i$ vectors respectively. Of course, we can freely choose either $\chi_i$ or ${\mathcal L}_i$. However, if we do it in terms of ${\mathcal L}_i$ instead of $\chi_i$, it leaves us with an unwanted remnant freedom in the system.

Altogether the 12 variables, defined in the equations (\ref{tetrpert}) and left alive after implying the diffeomorphism-gauge fixing (\ref{ourgauge}), are classified into 3 dynamical ($h_{ij}$, $s$), 6 constrained ($\phi$, $\psi$, ${\mathcal M}_i$, ${\mathcal L}_i$), and 3 gauge freedoms ($\zeta$, $\chi_i$). Note though that the purely Lorentzian modes are not directly observable anyway since they don't influence the metric. As we mentioned above, for many people it was a clear motivation for preferring precisely such models. However, there are also serious arguments \cite{Jose} that it might be impossible to keep the extra gauge freedom unbroken in the non-linear corrections.

\subsubsection{The $a=b\neq 0$, $a\neq c$, $a\neq 3c$ model}

This model, of restriction II, keeps $a=b$ as in GR but changes the spherically symmetric solutions. The perturbations behave as
\begin{equation*}
\left\{
\begin{array}{l}
{\ddot\zeta} - \bigtriangleup\zeta =0,\\
\phi=-\psi=\frac{c-a}{3c-a}\cdot \dot\zeta,\\
{\mathcal M}_i =0,\\
\bigtriangleup\chi_i=-{\dot{\mathcal L}}_i,\\
{\ddot h}_{ij} -\bigtriangleup h_{ij}=0.
\end{array}
\right.
\end{equation*}
We can say that the 12 variables come as 3 dynamical ($h_{ij}$, $\zeta$), 6 constrained ($\phi$, $\psi$, ${\mathcal M}_i$, $\chi_i$), and 3 gauge-free ($s$, ${\mathcal L}_i$). The pair of $\chi_i$ and  ${\mathcal L}_i$ is classified into constrained and gauge-free in such a way as to maximally avoid any remnant freedom. Note that the Lorentz boost waves $\zeta$ are now observable in the metric (conformal mode). The same as in the case above, this model can hopefully be taken as a well-defined one as long as this dynamical structure is robust against the higher-order corrections.

\subsubsection{The $a=-b\neq 0$, $c\neq 0$ model}

Now the codimension-one cases go far beyond the GR. The restriction III case yields
\begin{equation*}
\left\{
\begin{array}{l}
 \bigtriangleup\phi=-3{\ddot\psi} + 2 \bigtriangleup\psi,\\
\bigtriangleup\zeta=-3\dot\psi,\\
{\ddot s} - \bigtriangleup s =0,\\
 {\dot{\mathcal M}}_i = {\ddot\chi}_i - \bigtriangleup \chi_i,\\
{\mathcal L}_i=- {\dot\chi}_i.
\end{array}
\right.
\end{equation*}
The metric has completely lost any predictability in both its vector and tensor sectors, and partially in the scalar sector, too. We have 2  dynamical modes ($2=1+2\times\frac12$ as $s$ and one Cauchy datum for ${\mathcal M}_i$), 5 constrained variables ($5=4+2\times\frac12$ as $\phi$, $\zeta$, ${\mathcal L}_i$ plus half freedom killed for ${\mathcal M}_i$), and 5 gauge freedoms ($\psi$, $\chi_i$, $h_{ij}$).

Note that we refused to call $\psi$ a dynamical mode despite its acceleration term in the equations. It also appears in GR. The only difference here is that, instead of $\phi=\psi=0$, the constraint structure is much weaker. One could indeed use the gauge freedom of this model for choosing $\phi$, and then $\psi$ would have a dynamical equation. This is nothing but incomplete gauge fixing. Analogously, the $U(1)$ symmetry $A_{\mu}\longrightarrow A_{\mu}+\partial_{\mu}\lambda$ of electrodynamics can be fixed by $A_0=0$ with the remnant freedom of $\dot\lambda=0$. Or one could even gauge-fix it demanding $\square A_0=0$. It's definitely possible but leaves even a larger remnant freedom of $\square\dot\lambda=0$.

\subsubsection{The $6c=a+b$, $|a|\neq|b|$ model}

The selected restriction is IV. This is yet another far-from-GR case which corresponds to vanishing kinetic energy of the conformal mode. We get
\begin{equation*}
\left\{
\begin{array}{l}
\zeta =0,\\
\phi=-\psi,\\
{\ddot s} - \bigtriangleup s =0,\\
{\ddot{\mathcal M}}_i- \bigtriangleup{\mathcal M}_i =0,\\
{\ddot\chi}_i - \bigtriangleup \chi_i = -\frac{3b-a}{b-a}\cdot {\dot{\mathcal M}}_i,\\
{\mathcal L}_i=\frac{a+b}{a-b}\cdot {\mathcal M}_i - {\dot\chi}_i,\\
{\ddot h}_{ij} -\bigtriangleup h_{ij}=0.
\end{array}
\right.
\end{equation*}
The conformal mode is unpredictable, however the vectorial metric perturbations become dynamical.  Altogether there are 7 dynamical modes (${\mathcal M}_i$,  $h_{ij}$, $s$, $\chi_i$), 4 constrained ones ($\phi+\psi$, $\zeta$, ${\mathcal L}_i$) and 1 gauge freedom ($\phi-\psi$).

\subsubsection{The $a=b=c\neq 0$ model}

Now we come to codimension-two cases of parameter choices, and start from GR, i.e. the restrictions I and II.
\begin{equation*}
\left\{
\begin{array}{l}
\phi=\psi=0,\\
{\mathcal M}_i =0,\\
{\ddot h}_{ij} -\bigtriangleup h_{ij}=0.
\end{array}
\right.
\end{equation*}
As usual, 2 dynamical and 4 constrained modes in the metric, and all 6 gauge freedoms in the Lorentz sector.

\subsubsection{The $a=b=3c\neq 0$ model}

The mildest non-GR codimension-two choice is applying the restrictions II and IV:
\begin{equation*}
\left\{
\begin{array}{l}
\zeta =0,\\
\phi=-\psi,\\
{\mathcal M}_i =0,\\
\bigtriangleup\chi_i=-{\dot{\mathcal L}}_i,\\
{\ddot h}_{ij} -\bigtriangleup h_{ij}=0.
\end{array}
\right.
\end{equation*}
The conformal mode of the metric is unpredictable, though otherwise the metric sector is the usual one. And we see 2 dynamical modes (the standard GWs $h_{ij}$), 6 constrained ones ($\phi+\psi$, ${\mathcal M}_i$ $\zeta$, ${\chi_i}$), 4 gauge freedoms ($\phi-\psi$, $s$, ${\mathcal L}_i$).

\subsubsection{The $a=b=0$, $c\neq 0$ model}

Then the craziest models come. We have an option of the restrictions II and III with almost no information about the metric:
\begin{equation*}
\left\{
\begin{array}{l}
 \bigtriangleup\phi=-3{\ddot\psi} + 2 \bigtriangleup\psi,\\
\bigtriangleup\zeta=-3\dot\psi,\\
\bigtriangleup \chi_i = - {\dot{\mathcal M}}_i - {\dot{\mathcal L}}_i.
\end{array}
\right.
\end{equation*}
There are 0 dynamical modes, 4 constrained variables ($\phi$, $\zeta$, $\chi_i$), and 8 gauge freedoms ($\psi$, ${\mathcal M}_i$, $h_{ij}$, $s$, ${\mathcal L}_i$).

\subsubsection{The $a=-b\neq 0$, $c=0$ model}

The final option is the one with the restrictions I and III and IV. Despite having three restrictions, it is still a codimension-two case since any two of them immediately imply the third one.
\begin{equation*}
\left\{
\begin{array}{l}
{\ddot s} - \bigtriangleup s =0,\\
{\mathcal L}_i=- {\dot\chi}_i.
\end{array}
\right.
\end{equation*}
We get just 1 dynamical mode ($s$), 2 constraints (fixing ${\mathcal L}_i$ in terms of $\chi_i$), and 9 gauge freedoms including absolutely all the directly observable stuff (the metric).

\section{Conclusions}

We have presented a thorough analysis of the weak field limit (around the trivial, unit-matrix tetrad) for all possible classes of New General Relativity. In terms of the kinetic matrix, it fully reproduces the previouis results on primary constraints of these models. Actually, it must have been so, simply because the whole action of New GR is quadratic in velocities, by its very definition. However, the dynamical structure is not only about primary constraints, and the secondary ones were not known. Of course, what we have done is only a linearised dynamics around one particular background. However, even despite it being an extremely simple background, it is also a very interesting one and constitutes an important step towards understanding these models.

Our conclusion is that generic models with no extra restriction on the parameter space are the most promising ones. They have 8 dynamical degrees of freedom, precisely what is to be expected from the sixteen available variables after the four diffeomorphism gauge symmetries have "hit it twice". Most probably, it is the same in the non-linear regime, too. And even if something breaks down around some other backgrounds, it should be happening on some singular surfaces in the phase space, so that non-linear perturbative corrections are not expected to spoil any nice qualities of the linear weak field model.

All in all, it means that subsequent Hamiltonian analyses of New GR should focus on generic values of the model parameters. Most other versions do not even look like gravity theories. The only two other non-GR options which also have some promise are those which are obtained by imposing one of the two restrictions which, when taken together, reduce the model to GR. Any one of these choices brings an extra gauge symmetry for one half of the Lorentz sector. Therefore, another important question to future works is whether this gauge freedom can be kept stable in the full theory without permeating the physical metric geometry.

\end{document}